# On the Superconductivity of $LaFe_{1-y}Co_yAsO_{1-x}F_x$

Ayaka Kawabata, Sang Chul Lee, Taketo Moyoshi, Yoshiaki Kobayashi and Masatoshi Sato*

*Department of Physics, Division of Material Science, Nagoya University, Furo-cho, Chikusa-ku, Nagoya 464-8602*

We have prepared the superconducting system $LaFe_{1-y}Co_yAsO_{1-x}F_x$ ($x$=0.11) and carried out measurements of their electrical resistivities ρ and superconducting diamagnetisms. $^{75}$As- and $^{135}$La-NMR studies have also been carried out. The Knight shift observed for $^{75}$As has been found to be suppressed by the superconductivity, while for $^{135}$La, the shift is almost insensitive to the superconductivity. This result presents rather strong experimental evidence for the singlet pairing. The Co-doping effect on the superconducting transition temperature $T_c$ is not so significant as expected for superconductors with nodes, suggesting that the potential scattering does not seem to primarily suppress the superconductivity. Even for superconductors without nodes, it may not be so trivial to expect this small effect, if there are two different (disconnected) Fermi surfaces whose order parameters have opposite signs. As a possible explanation of the observed $T_c$ suppression, which is found to be small, seems to be related with the loss of the itinerant nature of the electrons..

*Corresponding author: e43247a@nucc.cc.nagoya-u.ac.jp

The discovery of the Fe pnictite superconductor $LaFeAsO_{1-x}F_x$ with the transition temperature $T_c$ ~ 26 K[1)] has presented a remarkable example of 3$d$ electron superconductors, which follows Cu oxides[2)] and $Na_{0.3}CoO_2 \cdot 1.3H_2O$.[3)] The system attracts much attention because it has strongly correlated electrons and also because many related systems can be derived by the substitution of the constituent elements. Actually, by the total substitution of La atoms with various other lanthanide elements Ln, the superconductivity with $T_c$ higher than 50 K has been reported.[4)] The superconducting transition with $T_c$ ~ 38 K has also been reported for $Ba_{1-x}K_xFe_2As_2$.[5)]

These systems commonly have FeAs layers formed of edge-sharing $FeAs_4$ tetrahedra,[1)] and the superconductivity is considered to primarily take place in these FeAs layers,[6, 7)] that is, the 3$d$-electrons usually expected to be strongly correlated, exhibit the high $T_c$ values. The spin-density-wave (SDW)-like transition observed for the mother system LaFeAsO[8)] suggests that the magnetic interaction cannot be ignored in the study of this superconductivity.

We have synthesized $LaFe_{1-y}Co_yAsO_{1-x}F_x$ ($x$=0.11) and used for measurements of electrical resistivities ρ and magnetizations $M_s$ due to the superconducting diamagnetism, where Co-doping effects on the superconductivity of this new system has mainly investigated, expecting that results of the studies can present us wealthy information on various properties of the system, such as the electronic state and origin of the superconductivity. We have also carried out $^{75}$As- and $^{135}$La-NMR measurements to study the microscopic nature of the systems. On the basis of these measurements, we show that the superconducting pairs are in the singlet state. We also show that the $T_c$-suppression by doped Co-impurities is rather weak as compared with that observed for Cu oxides,[9)] and argue what the results imply on the superconducting state.

Polycrystalline samples of $LaFe_{1-y}Co_yAsO_{1-x}F_x$ ($x$=0.11) were prepared from initial mixtures of La, $La_2O_3$, $LaF_3$ and FeAs with the nominal molar ratios. They are ground in a glove box filled with Ar gas and pelletized. Then, the pellets were sealed in an evacuated quartz tube and slowly heated up. Finally they were heated at 1150°C for ~38 h, slowly cooled to 700°C and then, furnace-cooled.

By X-ray measurements with Cu$K\alpha$ radiation, we have often observed LaOF as an impurity phase, whose typical molar concentration ~0.03. For some samples, there has been found a small amount of the ferromagnetic Fe phase, whose molar fraction is estimated to be less than $2.5\times10^{-3}$. We can safely say that the Co doping is really achieved here, because we have observed the change of the lattice parameter $c$ consistent with the linear dependence on $y$ studied in the wide region $0\le y \le 1$,[9, 10)] though we could not find a meaningful change of the lattice constant with $y$ in the region $y\le$ 0.03 due to the smallness of their expected fractional change (≤0.001).

The superconducting diamagnetism was measured by a Quantum Design SQUID magnetometer with the magnetic field of 10 G under both the zero-field-cooling (ZFC) and field-cooling (FC) conditions. The electrical resistivity ρ was measured by the four-terminal method with increasing $T$. The $T_c$ values are defined as the crossing temperatures of the linear extrapolation of the $M_s$-$T$ curve from the $T$ region below $T_c$ with the normal state magnetization extrapolated from the region above $T_c$. These $T_c$ values are in good agreement with those estimated by the extrapolation of the ρ-$T$ curve to ρ = 0.

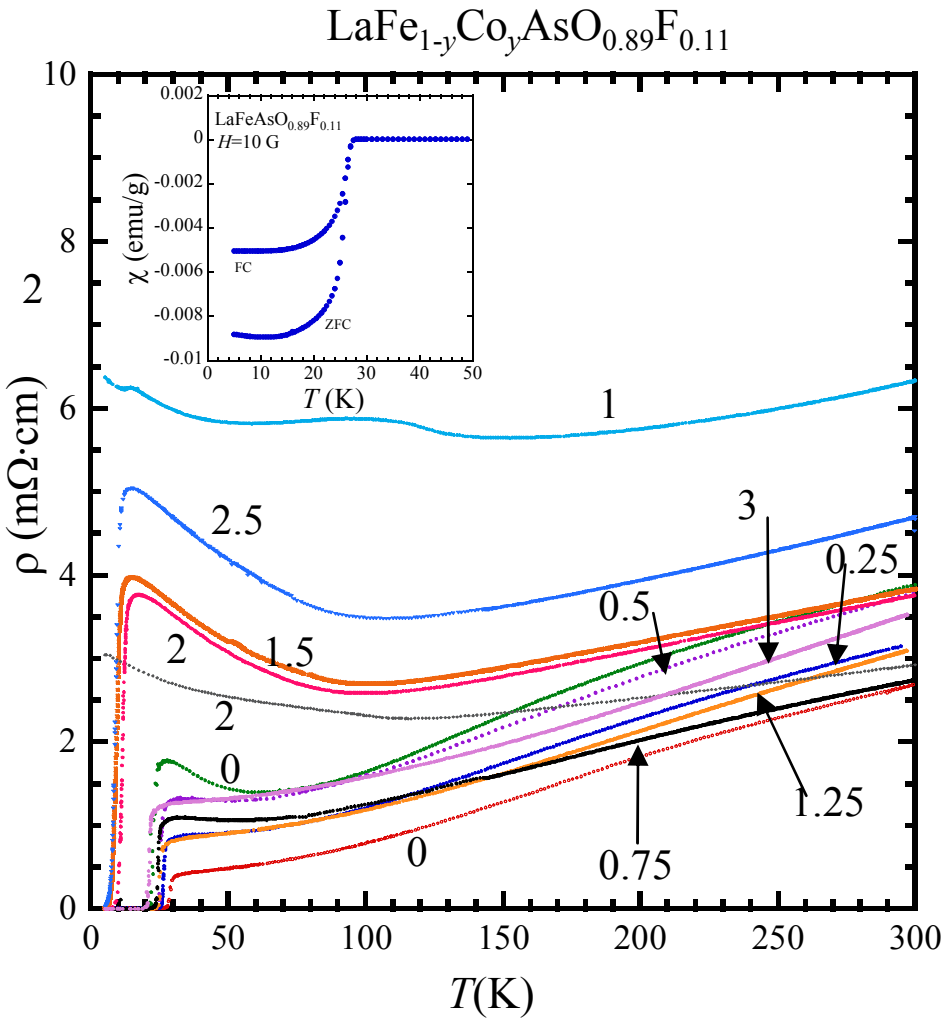

Fig. 1 Temperature dependence of the electrical resistivities of polycrystalline samples of $LaFe_{1-y}Co_yAsO_{1-x}F_x$ ($x$=0.11) The values of 100$y$ are attached to the corresponding curves. The inset shows a typical example of the superconducting diamagnetism observed with the FC and ZFC conditions.

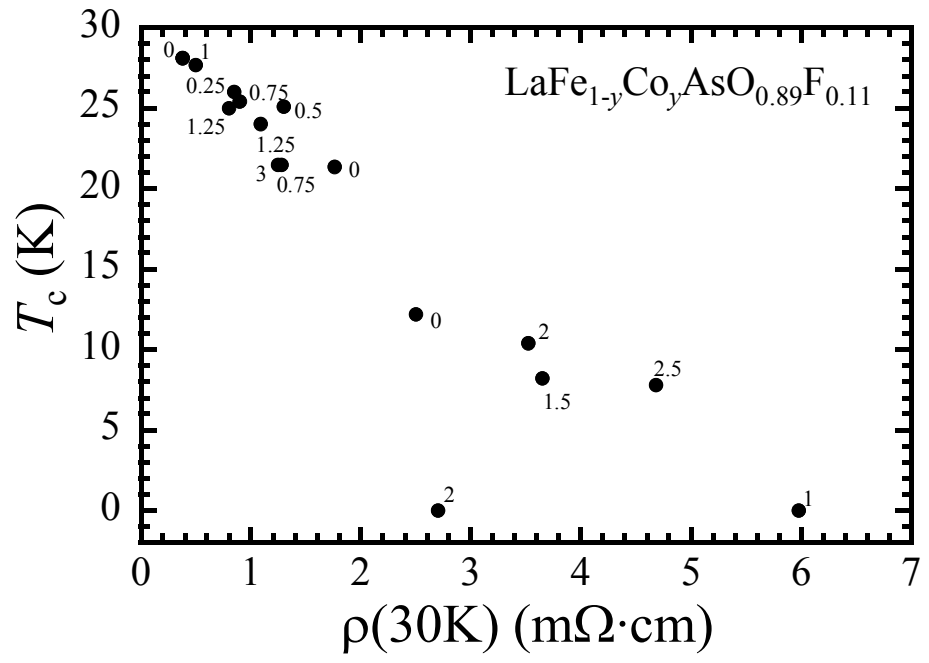

Fig. 2 $T_c$ values determined from the $T$ dependence of $M_s$ are shown against the ρ values at 30 K. The values of 100$y$ are attached to the corresponding data.

The $^{75}$As- and $^{135}$La-NMR measurements were carried out by the standard coherent pulse method. The $^{75}$As- and $^{135}$La-NMR spectra were measured by recording the nuclear spin-echo intensity $I$ with the NMR frequency or applied magnetic field being changed stepwise.

Figure 1 shows the electrical resistivities against $T$ for the samples with various nominal $y$ values. Typical data of $M_s$ are shown in the inset, where the volume fraction is estimated to be ~80% for the ZFC condition. The $T_c$ values estimated from resistivity curves do not seem to have meaningful correlation with $x$. Instead, a tendency exists that the higher $T_c$ values are found for the samples with the smaller resistivity below 150 K. To show this correlation between $T_c$ and ρ more directly, we plot the $T_c$ values estimated from the $M_s$ data against ρ at 30 K, ρ(30 K), in Fig. 2, where the 100$y$ values are attached to the corresponding points. We stress again that we can hardly find meaningful correlation between $T_c$ and $y$, which imply that, at least, the carrier scattering by Co impurities is not playing a primary role in the $T_c$ suppression. If the observed suppression were mainly due to the potential scattering of the carrier electrons by Co impurities, $T_c$ had to have a clear correlation with $y$. Therefore, we think that the suppression of $T_c$ due to the carrier scattering by Co impurities is so small that the complete suppression is not achieved by the Co doping of several percents. This result is rather important, because for superconductors with node(s) of the order parameter, $T_c$ is rapidly suppressed by the potential scattering of the carrier electrons. Even for superconductors without nodes, the observed small suppression rate of $T_c$ may not be easily expected, if there are two different (disconnected) Fermi surfaces whose order parameters have opposite signs. This adds further information to the nodeless nature[12-15] of the superconductivity of the present system. The small suppression rate of $T_c$ by the impurity scattering seems to be consistent with the reported appearance of the superconducting transitions in $LaFe_{1-y}Co_yAsO$ and $AFe_{2-y}Co_yAs_2$ (A=Ba, Sr) with $y$ much larger than those of the present samples.[10, 11, 16, 17]

The relatively rapid drop of $T_c$ is found in Fig. 2 in the ρ(30 K) region between 2-3 mΩ·cm. Although the intrinsic ρ cannot be precisely estimated for the polycrystalline samples because of the possible grain-boundary scattering, we speculate that the loss of the itinerant nature of the electrons is a possible origin of the $T_c$ decrease. For this kind of two-dimensional conductors, the sheet resistance $R_\square$ of 6.45 kΩ is a metal-insulator phase boundary,[9] which corresponds to a much smaller value (~ 0.45 mΩ·cm ) of the grain-boundary-free ρ of the system. The speculation is supported by the log$T$ dependence of the resistivity observed by Riggs *et al.* for $SmFeAsO_{1-x}F_x$ under the field above the critical field $H_{c2}$.[18]

Experimental data, which suggest that the $T_c$ decrease is induced by the increase of ρ, can be found in the present NMR results, too. In Fig. 3, we plot the logarithm of the integrated spin-echo intensities $I$ of $^{139}$La against 2τ, τ being the time between the first and second pulses. They were obtained at the fixed field $H$ ~ 7.4 T. (In the figure, we use [$I$(2τ)/$I$(2τ=20μs)]) instead of $I$(2τ).) While the curves are linear for all samples at high temperatures (top panel), they become nonlinear with decreasing $T$. This deviation from the linear line is more significant for the samples with smaller $T_c$ values, indicating that the magnetic fluctuation effect on the nuclear spin-spin relaxation is larger in the samples with lower $T_c$ or larger ρ (bottom panel of Fig. 3).

As can be expected from the results shown in Fig. 3, we have observed the so-called wipeout, which is well-known as a phenomenon that the NMR intensity decreases or vanishes for certain reasons: For example, in the top panel of Fig. 4, we plot the $^{139}$La-NMR intensities multiplied by $T$, ($I$×$T$), for the sample with $y$=0.0 ($T_c \sim 12$ K) at two temperatures against $H$. They were measured with $2\tau \sim 60\ \mu s$ at the fixed frequency $f$ = 44.6 MHz. ($I \times T$) clearly decreases with decreasing $T$. (In this case, its origin is the rapid decay of the signal with $2\tau$ shown in Fig. 3.)

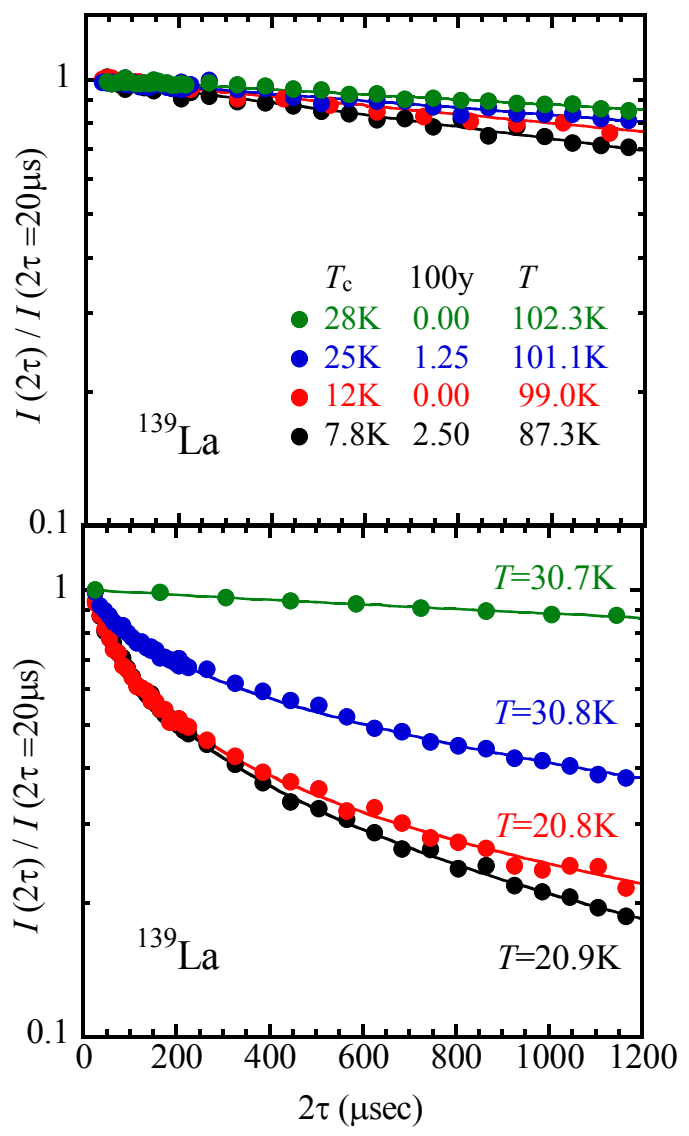

Fig. 3 Integrated NMR intensities $I$ divided by $I(2\tau = 20\ \mu s)$ are shown against $2\tau$. They were obtained at the fixed field $H \sim 7.4$ T. At high temperatures, all the data show the Gaussian-like $2\tau$ dependence (top). At low temperatures, the curves except the one obtained for the sample with the highest $T_c$ deviate from the Gaussian-like $2\tau$ dependence, indicating that the spin-spin relaxation due to the strong magnetic fluctuation becomes significant (bottom). Note that the deviation is larger for samples with lower $T_c$.

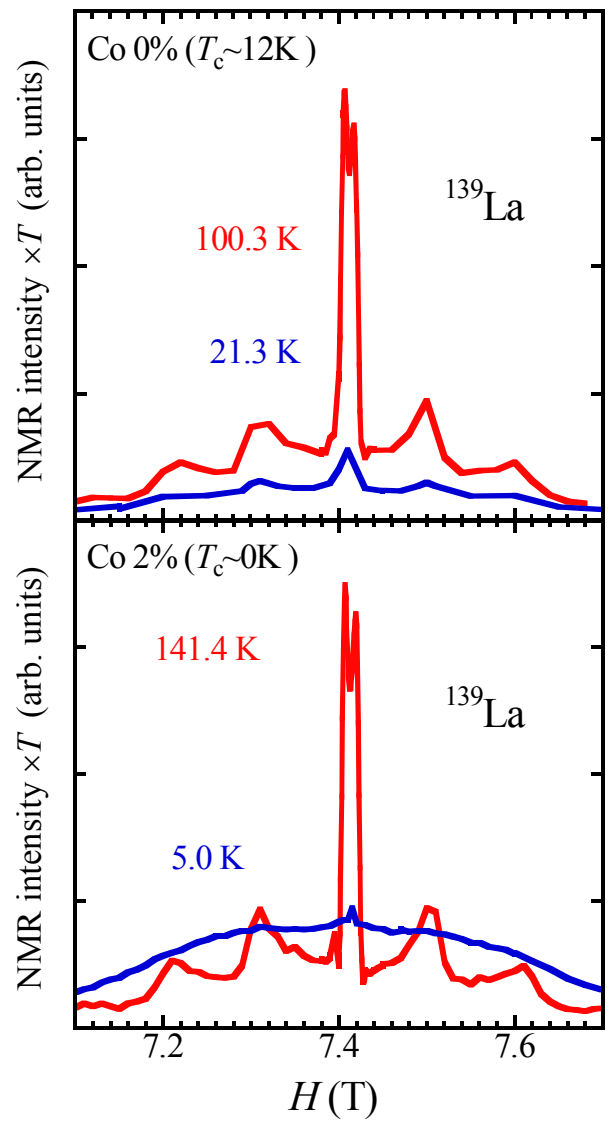

Fig. 4 Spin-echo intensities of $^{139}$La multiplied by $T$ are shown against the magnetic field $H$ ($f$ = 44.6 MHz). For the sample with $T_c \sim 12$ K, the wipeout effect can be observed (top). For the sample which does not exhibit superconductivity down to 5 K, the significant broadening of the spectra can be seen at 5 K, indicating the static antiferromagnetic ordering exists.

The data in Figs. 2-4 indicate that as the system approaches the metal-insulator boundary, or as the loss of the itinerant nature of the electrons becomes significant, the magnetic fluctuation becomes stronger and $T_c$ decreases. For the sample with $T_c$ <5 K and $y$= 0.02, whose ρ-$T$ curve can be found in Fig. 2, we have observed the large broadening of the $^{139}$La-NMR spectra at 5 K (bottom panel of Fig. 4), indicating that the magnetically ordered phase appears near the superconducting phase. From the detailed $T$ dependence of its $\log[I(2\tau)/I(2\tau=20\mu s)]$-$2\tau$ curve, we have found that the antiferromagnetic (or SDW) ordering takes place at ~ 60 K. We have not observed significant anomaly in the $T$ dependences of ρ and the magnetic susceptibility.

In the left column of Fig. 5, the $^{139}$La- and $^{75}$As -NMR spectra measured at $H$= 6.05 T for a sample with $y$=0 and $T_c \cong 28$ K in zero field, are shown against the NMR frequency $f$, where the main panels show the data around the positions indicated by the arrows in the insets. It can be clearly found that with decreasing $T$, the peak positions of $^{139}$La and $^{75}$As begin to exhibit significant shifts toward the opposite sides at $T_c$, and the shift magnitude is much larger for $^{75}$As than that for $^{139}$La. The deviations $\Delta f$ of the peak positions from those at 40 K are shown against $T$ in the upper right panel of Fig 5, and they are re-plotted in the lower right panel, in the form of the deviations of Knight shift, $\Delta K$. It is quite apparent that only for $^{75}$As, $\Delta f$ is significant, indicating that it does not originate from the superconducting diamagnetism (If the diamagnetic field were important, the shifts of $^{139}$La and $^{75}$As would have a same sign.). The slight $T$ dependence above $T_c$ can be understood as the pseudogap-like behavior reported in ref. 19.

These results indicate that the Cooper pairs are in the singlet state. The $T$ dependence of $\Delta K$ can roughly be explained by the Yosida function (broken lines) for the $s$-symmetry order parameter,[20] although it may not be precise enough to restrict that the order parameter obeys the simple $s$-wave form. However, it does not exhibit the unusual behavior reported by Matano *et al.*,[21] which suggests the existence of two gaps.

We have shown the electrical resistivity ρ and superconducting diamagnetism $M_s$ of $LaFe_{1-y}Co_yAsO_{1-x}F_x$ ($x$=0.11). The results of $^{75}$As- and $^{135}$La-NMR studies have also been shown. The suppression of $K$ presents a strong experimental evidence for the singlet pairing. The potential scattering by doped impurities does not seem to suppress the superconductivity so significantly as compared with those of high $T_c$ Cu oxides.[9] This small effect on $T_c$ cannot usually be expected for the superconductors with nodes. It may not be so trivial to expect this small effect even for superconductors without nodes, if there are two different (disconnected) Fermi surfaces whose order parameters have opposite signs. This result presents a new important information on the

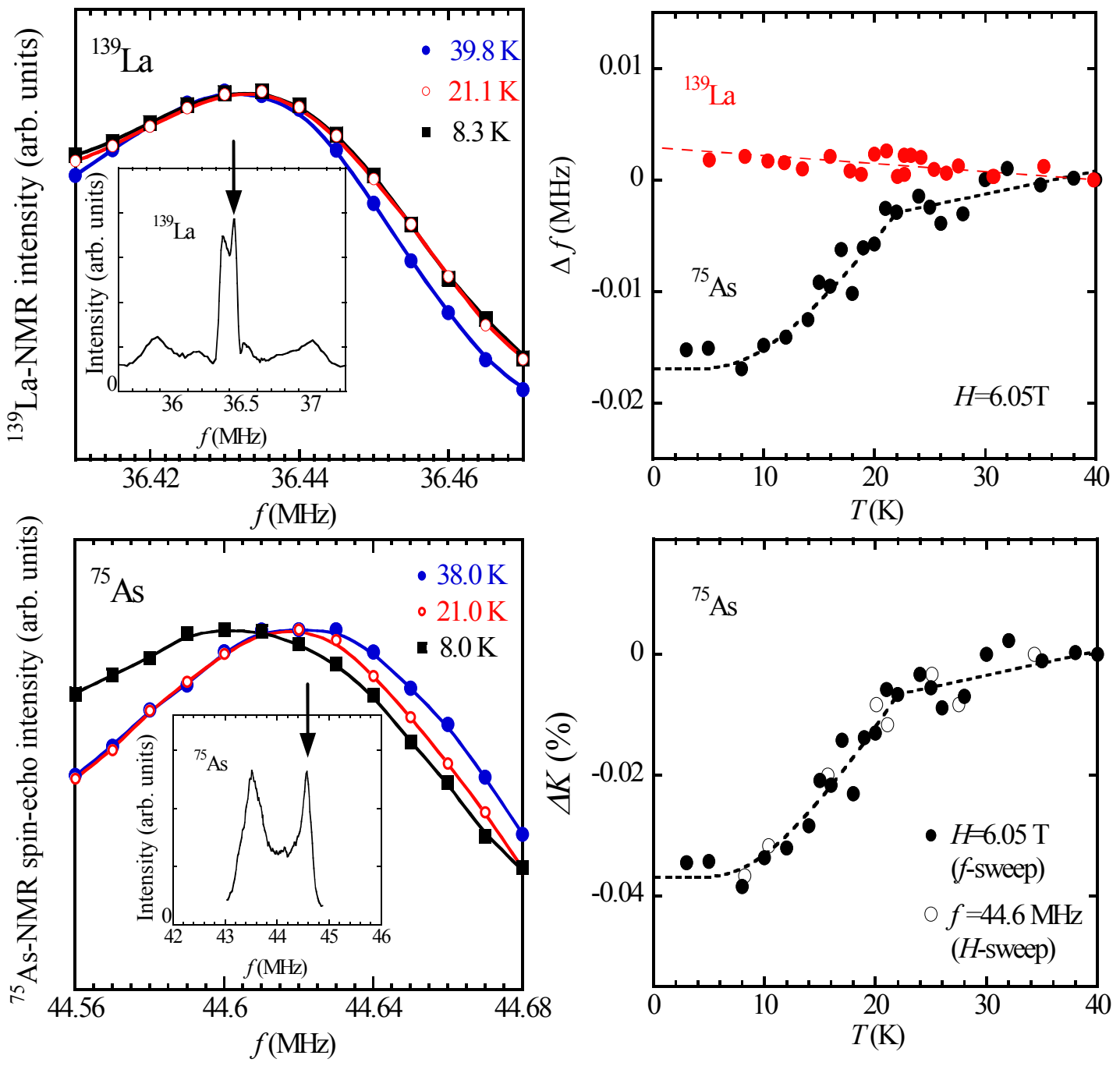

Fig. 5 The left column shows the $^{139}$La and $^{75}$As NMR intensities against the NMR frequency $f$. They were obtained for a sample with $y$=0 and $T_c \cong$ 28 K at $H$= 0: The main panels show the data around the positions indicated by the arrows in the insets. The deviations $\Delta f$ of the peak positions from those at 40 K are shown against $T$ in the upper right panel of Fig 5, and they are re-plotted in the lower right panel in the form of the $\Delta K - T$ curves.

issue whether nodes of the gap parameter exists, because many conflicting data have been reported by measuring NMR longitudinal relaxation rate $1/T_1$,[19, 21] magnetic field penetration depth,[12, 13] and photoemission spectroscopy.[14, 15]

A possibility has been pointed out that the $T_c$ suppression, which is found to be rather small, seems to be related with the loss of the itinerant nature of the electrons.

Acknowledgments –The work is supported by Grants-in-Aid for Scientific Research from the Japan Society for the Promotion of Science (JSPS) and by Grants-in-Aid on Priority Area from the Ministry of Education, Culture, Sports, Science and Technology.

1) Y. Kamihara, T. Watanabe, M. Hirano, and H. Hosono; J. Am. Chem. Soc. **130** (2008) 3296.
2) J. G. Bednorz, and K. A. Muller; Z. Phys. **B64** (1986) 189.
3) K. Takada, H. Sakurai, E. Takayama-Muromachi, F. Izumi, R. A. Dilanian, and T. Sasaki, Nature **422** (2003) 53.
4) Zhi-An Ren, Jie Yang, Wei Lu, Wei Yi, Xiao-Li Shen, Zheng-Cai Li, Guang-Can Che, Xiao-Li Dong, Li-Ling Sun, Fang Zhou and Zhong-Xian Zhao; Europhys. Lett. **82** (2008) 57002.
5) M. Rotter, M. Tegel, and D. Johrendt: arXiv:0805.4630.
6) S. Ishibashi, K. Terakura, and H. Hosono; J. Phys. Soc. Jpn. **77** (2008) 053709.
7) K. Haule, J. H. Shim, and G. Kotliar; Phys. Rev. Lett. **100** (2008) 226402.
8) Clarina de la Cruz, Q. Huang, J. W. Lynn, Jiying Li, W. Ratcliff II, J. L. Zarestky, H. A. Mook, G. F. Chen, J. L. Luo, N. L. Wang, Pengcheng Dai; Nature **453** (2008) 899 - 902.
9) H. Harashina, T. Nishikawa, T. Kiyokura, S. Shamoto, M. Sato and K. Kakurai: Physica C **212** (1993) 142.
10) A. S. Sefat, A. Huq, M. A. McGuire, R.Jin, B. C. Sales, D. Mandrus; arXiv: 0807.0823.
11) G. Cao, C. Wang, Z. Zhu, S. Jiang, Y. Luo, S. Chi, Z. Ren, Q. Tao,, Y. Wang, Z. Xu; arXiv: 0807.1304.
12) K. Hashimoto, T. Shibauchi, T. Kato, K. Ikada, R. Okazaki, H. Shishido, M. Ishikado, H. Kito, A. Iyo, H. Eisaki, S. Shamoto, Y. Matsuda; arXiv:0806.3149v2.
13) L. Malone, J.D. Fletcher, A. Serafin, A. Carrington, N.D. Zhigadlo, Z. Bukowski, S. Katrych, J. Karpinski; arXiv:0806.3908v1.
14) T. Kondo, A. F. Santander-Syro, O. Copie, C. Liu, M. E. Tillman, E. D. Mun, J. Schmalian, S. L. Bud'ko, M. A. Tanatar, P. C. Canfield, and A. Kaminski; arXiv:0807.0815.
15) L. Zhao, H. Liu, W. Zhang, J. Meng, X. Jia, G. Liu, X. Dong, G. F. Chen, J. L. Luo, N. L. Wang, G. Wang, Y. Zhou, Y. Zhu, X. Wang, Z. Zhao, Z. Xu, C. Chen, X. J. Zhou; arXiv:0807.0398.
16) A. S. Sefat, A. M. A Mcguire, R. Jin, B. C. Sales, D. Mandrus; arXiv: 0807.2237.
17) A. Leithe-Jasper, W. Schnelle, C. Geibel, and H. Rosner; arXiv: 0807.2223v1.
18) S. C. Riggs, J. B. Kemper, Y. Jo, Z. Stegen, L. Balicas, G. S. Boebinger, F. F. Balakirev, A. Migliori, H. Chen, R. H. Liu, and X. H. Chen; arXiv: 0806.4011v1.
19) Y. Nakai, K. Ishida, Y. Kamihara, M. Hirano, and H. Hosono; J. Phys. Soc. Jpn. **77** (2008) 073701-(1-4).
20) K. Yosida: Phys. Rev. **110** (1958) 769.
21) K. Matano, Z. A. Ren, X. L. Dong, L. L. Sun, Z. X. Zhao, and G.-q. Zheng; arXiv: 0806.0249.